# Effect of inhomogeneous Dzyaloshinskii-Moriya interaction on antiferromagnetic spin-wave propagation

Seung-Jae Lee,[1] Dong-Kyu Lee,[2] and Kyung-Jin Lee[1,2*]

[1]KU-KIST Graduate School of Converging Science and Technology, Korea University, Seoul 02841, Korea

[2]Department of Materials Science and Engineering, Korea University, Seoul 02841, Korea

* Corresponding Email: kj_lee@korea.ac.kr

Abstract

We investigate the effect of inhomogeneous Dzyaloshinskii-Moriya interaction (DMI) on antiferromagnetic spin-wave propagation theoretically and numerically. We find that antiferromagnetic spin waves can be amplified at a boundary where the DMI varies. The inhomogeneous DMI also provides a way to construct a magnonic crystal with forbidden and allowed antiferromagnetic spin-wave bands in terahertz frequency ranges. In contrast to ferromagnetic spin waves, antiferromagnetic spin waves experience a polarization-dependent phase shift when passing through the inhomogeneous DMI, offering a magnonic crystal that also serves as a spin-wave polarizer.

I. Introduction

Recently, much effort has been expended in employing antiferromagnets for spintronics research [1,2]. An important motivation is that spin dynamics in antiferromagnets is much faster than in ferromagnets. Net zero spin density of two antiferromagnetically coupled sub-lattices suppresses the rotational motion of antiferromagnetic spin textures, which allows antiferromagnetic domain walls move much faster than ferromagnetic domain walls [3-11] and also results in vanishing skyrmion Hall effect in antiferromagnets [12,13] or compensated ferrimagnets [14]. Antiferromagnetic exchange interaction between neighboring spin moments provides a high resonance frequency in the terahertz (THz) ranges [15,16] leading to recent research efforts on THz spin oscillators [17-21].

Spin waves are low-energy collective excitations in magnetism. The ferromagnet has been considered as a core material for spin-wave devices where spin waves are used as information carriers [22-41]. As spin waves are able to deliver information without involving moving charges, they offer a computing scheme operated at low-power consumption [25,29,36,40]. Because the spin-wave resonance frequency of antiferromagnets can reach THz, which is much higher than that of ferromagnetic spin waves, it is expected that antiferromagnetic spin waves may allow fast operation of spin-wave devices [42,43]. Another distinct feature of antiferromagnetic spin waves is that both right-handed and left-handed spin waves are allowed in antiferromagnets. These energetically degenerate two spin-wave modes offer rich spin-wave physics and additional knobs to operate antiferromagnetic spin-wave devices.

When magnetic systems are subject to the structural inversion asymmetry and spin-orbit coupling, the anti-symmetric exchange energy, called Dzyaloshinskii-Moriya interaction (DMI), emerges [44,45]. The DMI causes the nonreciprocal spin-wave propagation [46-48], providing a way to quantify the strength of DMI [49-52]. The DMI also offers widespread applications in functional spin-wave devices, by enabling unidirectional caustic beams [53], spin-wave diodes [31], and spin-wave fibers [32,33]. When combined with antiferromagnets and thus the spin-wave polarization (i.e., left- and right-handed), the DMI makes it possible to realize spin-wave transistor [54] and spin-wave phase shifter and retarder [43].

We have recently investigated the role of inhomogeneous DMI in spin-wave propagation [55], domain wall motion [56], and magnetic skyrmion motion [57], which are all for ferromagnets. We found that a spatial inhomogeneity of DMI creates a potential landscape for spin waves and magnetic solitons, which in turn affects spin-wave propagation [55] and soliton motion [56,57] substantially. Moreover, the inhomogeneous DMI acts as an effective equilibrium dampinglike spin-orbit torque proportional to the DMI gradient [55,57], which tilts the equilibrium magnetization direction [55-58] and causes a large transverse motion of a magnetic skyrmion, similar to the skyrmion Hall effect [57].

These interesting consequences of the inhomogeneous DMI for ferromagnetic spin waves and solitons motivate us to investigate antiferromagnetic spin-wave propagation in the presence of inhomogeneous DMI, theoretically and numerically, in this work. We first derive a boundary condition for antiferromagnetic spin waves at a DMI step at which the DMI strength changes. Then we combine this boundary condition with antiferromagnetic spin-wave dispersion and show that a concerted action of antiferromagnetic spin waves with the inhomogeneous DMI offers a high-frequency magnonic crystal, which allows not only frequency-dependent but also polarization-dependent control of spin-wave propagation.

## II. Antiferromagnetic spin-wave propagation across a DMI step

We consider a one-dimensional antiferromagnetic spin chain along the *x* direction [Fig. 1(a)]. The Lagrangian energy density is expressed as [59-61]

$$\mathcal{L} = -s\dot{\mathbf{n}} \cdot (\mathbf{n} \times \mathbf{m}) - \mathcal{U}, \tag{1}$$

where $s = M_s/\gamma$ is the local spin density, $M_s$ is the magnitude of magnetization, $\gamma = 1.76 \times 10^{11} \text{T}^{-1}\text{s}^{-1}$ is the gyromagnetic ratio, $\mathbf{n} \equiv (\mathbf{S}_1 - \mathbf{S}_2)/2$ is the staggered order parameter, $\mathbf{m} \equiv (\mathbf{S}_1 + \mathbf{S}_2)/2$ is the magnetic order parameter, $\mathbf{S}_{1,2}$ are the unit vectors along the magnetic moment at each sub-lattice, and $\mathcal{U}$ is given as

$$\mathcal{U} = \frac{A}{2}(\nabla \mathbf{n})^2 + \frac{a}{2}(\mathbf{m})^2 - \frac{K}{2}(\mathbf{n} \cdot \hat{\boldsymbol{p}})^2 - \frac{D(x)}{2}\mathbf{n} \cdot \left(\hat{\boldsymbol{y}} \times \frac{\partial \mathbf{n}}{\partial x}\right) - \frac{1}{4}\frac{dD(x)}{dx}\mathbf{n} \cdot (\hat{\boldsymbol{y}} \times \mathbf{n}), \tag{2}$$

where $\hat{\boldsymbol{p}}$ is the easy-axis for uniaxial magnetic anisotropy, and *A, a, K,* and *D(x)* are the inhomogeneous exchange, homogeneous exchange, anisotropy, and position-dependent DMI, respectively. In the strong exchange limit (i.e., $|\mathbf{m}| \ll |\mathbf{n}|$), the equation of motion for spin dynamics without damping is simplified as

$$\mathbf{n} \times \frac{\partial^2 \mathbf{n}}{\partial t^2} = a^*\gamma^2 \boldsymbol{n} \times \left[J^*\nabla^2 \mathbf{n} + K^*(\mathbf{n} \cdot \hat{\boldsymbol{p}})\hat{\boldsymbol{p}} + D^*(x)\left(\hat{\boldsymbol{y}} \times \frac{\partial \mathbf{n}}{\partial x}\right) + \frac{1}{2}\frac{dD^*(x)}{dx}(\hat{\boldsymbol{y}} \times \mathbf{n})\right], \tag{3}$$

where $a^* = \frac{2a}{M_s} = -\frac{8A}{M_s d^2}$, $J^* = \frac{2A}{M_s}$, $K^* = \frac{2K}{M_s}$, $D^*(x) = \frac{2D(x)}{M_s}$, and $d$ is the atomic distance. The last term proportional to $\frac{dD^*(x)}{dx}$ describes an equilibrium dampinglike spin-orbit torque originating from the inhomogeneous DMI. We call this last term as an equilibrium dampinglike spin-orbit torque because it arises without a current injection and its symmetry resembles the symmetry of a non-equilibrium (electrically induced) dampinglike spin-orbit torque. The reason that the inhomogeneous DMI creates an equilibrium spin torque is as follows: (i) the DMI can be interpreted as an equilibrium spin current [62,63], which survives even after integrating over whole Fermi surface, due to the inversion symmetry breaking. (ii) It is well-known that a divergence of spin current is one of the sources for spin torques [64], satisfying the angular momentum conservation. Combining (i) with (ii), one finds that the inhomogeneous DMI creates a non-zero divergence of spin current in equilibrium and thus causes an equilibrium spin torque. For ferromagnetic cases, previous studies found that this term tilts the equilibrium magnetization at a DMI step [55-58], creates an energy barrier for domain wall and skyrmion [56,57], and amplifies the spin-wave amplitude [55].

For numerical simulations, we solve the atomistic Landau-Lifshitz-Gilbert equation [4,65], given as

$$\frac{\partial \boldsymbol{S}_i}{\partial t} = -\gamma \boldsymbol{S}_i \times \mathbf{H}_{eff,i} + \alpha \boldsymbol{S}_i \times \frac{\partial \boldsymbol{S}_i}{\partial t}, \quad (4)$$

where $\boldsymbol{S}_i$ is the unit vector along the magnetic moment at a site $i$, $\mathbf{H}_{eff,i} = -\frac{1}{M_s}\frac{\partial W}{\partial \boldsymbol{S}_i}$, and the magnetic energy density $W$ is given as

$$W = \frac{2A}{d^2}\sum_i \boldsymbol{S}_i \cdot \boldsymbol{S}_{i+1} - K\sum_i (\boldsymbol{S}_i \cdot \widehat{\boldsymbol{p}})^2 - \frac{2D(x)}{d}\sum_i \widehat{\boldsymbol{y}} \cdot (\boldsymbol{S}_i \times \boldsymbol{S}_{i+1}) - M_s\mu_0 \sum_i \boldsymbol{S}_i \cdot \boldsymbol{H}. \quad (5)$$

We use following parameters for numerical simulations: Atomic distance $d$ = 0.4 nm, exchange stiffness constant $A = -6.4$ pJ/m, easy-axis anisotropy constant $K = 0.1$ MJ/m$^3$, Gilbert damping constant $\alpha = 1\times10^{-4}$, and magnetic moment density $M_s$ = 400 kA/m. Spin waves are excited by applying circular magnetic fields with the magnitude of 1 mT.

### A. Easy axis aligned with y axis

We first consider the case for $\hat{p} = \hat{y}$, where the staggered order $\mathbf{n}$ is aligned along the $y$-axis in equilibrium. In this case, the last term in Eq. (3) vanishes and $\mathbf{n}$ is uniform in equilibrium. From Eq. (3) and using $\mathbf{n} = \mathbf{n}_0 + \delta\mathbf{n}$, where $\mathbf{n}_0$ is the position-independent staggered order, $\delta\mathbf{n} = (0, s_\theta, s_\phi)$ is the spin-wave contribution in the spherical coordinate ($s_\theta^2 \ll 1, s_\phi^2 \ll 1$), and $\theta$ and $\phi$ are the polar and azimuthal angles, respectively, we obtain the Schrödinger-like equation for spin-wave wave function $\psi \ (= s_\theta + is_\phi)$ for antiferromagnets, given as

$$\frac{\partial^2 \psi}{\partial t^2} = a^* \gamma^2 \left[ J^* \nabla^2 + K^* n_y + i n_y \left( D^*(x) \frac{\partial}{\partial x} + \frac{1}{2} \frac{\partial D^*(x)}{\partial x} \right) \right] \psi. \quad (6)$$

From the continuity of Eq. (6) and integrating Eq. (6) for $x$, we obtain the boundary conditions for $\psi$ at a DMI step [i.e., $D = D_1$ for $x < 0$ (region 1) and $D = D_2$ for $x \geq 0$ (region 2); see Fig. 1(a)], given as

$$\psi_1(x = 0) = \psi_2(x = 0), \quad (7)$$

$$\left.\frac{d\psi_1}{dx}\right|_{x=0} - \left.\frac{d\psi_2}{dx}\right|_{x=0} = \pm \Delta D \frac{in_y}{2A} \psi(x = 0), \quad (8)$$

where the positive (negative) sign corresponds to the right (left)-handed spin-wave mode, and $\Delta D \equiv D_2 - D_1$.

To verify the boundary conditions, we consider incident $\left(\psi_I = I_0 e^{ik_1 x},\right)$ and reflected $\left(\psi_R = A_0 e^{ik_2 x}\right)$ spin waves in the region 1, and transmitted $\left(\psi_T = B_0 e^{ik_3 x}\right)$ spin waves in the region 2, where $I_0$, $A_0$, and $B_0$ ($k_1$, $k_2$, and $k_3$) are the amplitude (wave vector) of incident, reflected, and transmitted waves. The spin-wave dispersion is given as

$$k^{(\pm)} = \mp \frac{D^*(x)}{2J^*} \pm \sqrt{\frac{K^*}{J^*} - \frac{\omega^2}{J^*\gamma^2 a^*} + \left(\frac{D^*(x)}{2J^*}\right)^2}, \qquad (9)$$

where $\hbar\omega$ is the spin-wave energy and the positive (negative) sign corresponds to the right (left)-handed spin-wave mode. From Eqs. (7)-(9), we obtain the analytic expression of spin-wave transmittance at a DMI step of antiferromagnets:

$$\frac{B_0}{I_0} = \frac{2\sqrt{D_1^{*2} + 4J^*H^*}}{\sqrt{D_1^{*2} + 4J^*H^*} + \sqrt{D_2^{*2} + 4J^*H^*}}, \qquad (10)$$

where $H^* = K^* - \omega^2/a^*\gamma^2$. Equation (10) predicts that the spin-wave transmittance does not depend on the spin-wave polarization (left- or right-handed) and spin waves can be amplified ($B_0/I_0 > 1$) at a DMI step when $|D_1^*| > |D_2^*|$. In particular, the spin-wave amplification becomes maximum when $D_2 = 0$.

In Fig. 1(b), we compare numerically-obtained spin-wave transmittances (symbols) with Eq. (10) (dotted lines) and find that they are in agreement, verifying the validity of the boundary conditions. As predicted by Eq. (10), the spin-wave transmittance is an even function of $D_2$ and the largest at $D_2 = 0$.

### B. Easy axis aligned with z axis

We next consider the case for $\hat{p} = \hat{z}$, where the staggered order **n** is aligned along the *z*-axis in equilibrium. In this case, the last term in Eq. (3) does not vanish and makes **n** tilt from the easy axis. Following Ref. [58], we obtain the equilibrium tilted texture for staggered order **n** in antiferromagnets, based on the Euler-Lagrange method. The equilibrium **n** texture in the present of a DMI step is given as

$$\mathbf{n} = (\sin\theta(x), 0, \cos\theta(x)), \tag{11}$$

where $\theta(x) = 2\arctan\left[e^{-|x/\xi|}\tan\frac{\theta_0}{2}\right]$, $\xi = \sqrt{-A/K}$, and the tilting angle at the DMI step is $\theta_0 = \arcsin\left[\frac{-\Delta D}{4\sqrt{-AK}}\right]$.

For propagating spin waves through non-uniform spin textures (e.g. magnetic domain walls), the spin wave acquires an additional phase due to the position-dependent effective field [34,51]. To calculate the phase shift in our case, we use the Wentzel-Kramers-Brillouin (WKB) approximation with the following rotation matrix **R**,

$$\mathbf{R}(\theta(x)) = \begin{bmatrix} \cos\theta(x) & 0 & -\sin\theta(x) \\ 0 & 1 & 0 \\ \sin\theta(x) & 0 & \cos\theta(x) \end{bmatrix}. \tag{12}$$

We transform the coordinate from $xyz$ to $\varepsilon\eta\zeta$ ($\eta = y$). In the transformed coordinate, the staggered order parameter is rewritten as $\mathbf{n} = \mathbf{n}_0 + \delta n_\varepsilon \hat{\boldsymbol{\varepsilon}} + \delta n_\eta \hat{\boldsymbol{\eta}}$, where $\mathbf{n}_0 = (0,0,1)$ and $\delta n_\varepsilon = s_\varepsilon \sin(k^\varepsilon x)$, $\delta n_\eta = s_\eta \sin(k^\eta x + \pi/2)$. Spin-wave dispersions for $\delta n_\varepsilon$ and $\delta n_\eta$, which are linear components for individual oscillating planes, are given as [43,66]

$$[k^\varepsilon(x)]^2 = \left(\frac{K^*}{J^*}\cos[2\theta(x)] - \frac{\omega^2}{J^*a\gamma^2}\right), \tag{13}$$

$$[k^\eta(x)]^2 = \left(\frac{K^*}{J^*}\cos[2\theta(x)] + \frac{D^*(x)}{J^*}\theta'(x) - \frac{\omega^2}{J^*a\gamma^2}\right). \tag{14}$$

Here we find that the two linear components ($\delta n_\varepsilon$, $\delta n_\eta$) have different wave vectors in the tilted region [i.e., $\theta'(x) \neq 0$]. On the other hand, in the uniform spin state [$x \to \pm\infty$ or $\theta'(x) = 0$], the wave vector of both linear components reduces to $\pm\sqrt{K^*/J^* - \omega^2/(J^*a^*\gamma^2)}$. This difference in the wave vector generates an additional phase shift for spin waves passing through a spin tilted region. The additional phase shift is calculated from the WKB approximation, given as

$$\Delta\phi_{x,y} = \int_{-\infty}^{\infty}(k^{\varepsilon,\eta}(x) - q)dx, \qquad (15)$$

where $q$ is the spin-wave wave vector in a region sufficiently far from the DMI step ($x \to \pm\infty$). Here we use subscripts $x$ and $y$, instead of $\varepsilon$ and $\eta$, for the phase shift $\Delta\phi$ because the phase shift is determined by comparing spin waves between far left and far right sides of the DMI step, where we can treat $k_\varepsilon$ ($k_\eta$) as $k_x$ ($k_y$).

Figure 2 shows the spin-wave phase shift originating from a DMI step. Here we excite spin waves with a left-handed circular polarization, which can be decomposed to two linear polarizations (i.e., $x$-linear and $y$-linear polarizations). Figure 2(a) shows $\Delta\phi_x$ and $\Delta\phi_y$ as a function of $D_2$ when $D_1 = -1$ mJ/m$^2$ and frequency $f = 0.6$ THz. For $D_2 = -1$ mJ/m$^2$, $\Delta\phi_x - \Delta\phi_y$ is 0 and thus there is no additional phase shift, because the DMI is homogeneous. For $D_2 = +0.8$ mJ/m$^2$, $\Delta\phi_x - \Delta\phi_y$ is $\pi/2$ so that the initially circularly-polarized spin waves become linear one after passing through the DMI step. In Fig. 2(a, b), we compare numerical phase shifts (symbols) with analytical ones (lines) and find that they are in reasonable agreement.

## III. Antiferromagnetic magnonic crystals with inhomogeneous DMI

Next, we consider a one-dimensional magnonic crystal composed of antiferromagnetic spin chain with alternating DMI ($D_1$ and $D_2$). We assume that other magnetic properties are uniform for simplicity. The spatial distribution of $D(x)$ is assumed to be

$$D(x) = \begin{cases} D_1, & nl < x < (n+1/2)l, \\ D_2, & (n+1/2)l < x < (n+1)l, \end{cases} \quad n = 0, 1, 2 ..., \qquad (16)$$

where $l/2$ is the width of a region with a homogeneous DMI. This alternating DMI serves as a periodic potential for spin waves and creates allowed and forbidden spin-wave bands.

In this section, we show how the spin-wave band structure can be engineered by inhomogeneous DMI in antiferromagnets. As the DMI is related with the inversion symmetry breaking and spin-orbit coupling, the spatial DMI strength can be modulated by either external electric fields [67-70] or locally varying the layer thickness [71-74].

### A. Easy axis aligned with y axis

Using the boundary conditions shown in Sec. II. A. and the Bloch's theorem, we obtain following spin-wave dispersion relation in a magnonic crystal with spatially modulated DMI constants ($D_1$ and $D_2$).

$$\cos\left[\tfrac{1}{4}l\left(4k + \tfrac{D_1^* + D_2^*}{J^*}\right)\right] = \cos\left[\tfrac{1}{2}l\mu\right]\cos\left[\tfrac{1}{2}lv\right] - \tfrac{(\mu^2 + v^2)}{2\mu v}\sin\left[\tfrac{1}{2}l\mu\right]\sin\left[\tfrac{1}{2}lv\right], \qquad (17)$$

where $\mu = \sqrt{\left(\tfrac{D_1^*}{2J^*}\right)^2 + \tfrac{H^*}{J^*}}$ and $v = \sqrt{\left(\tfrac{D_2^*}{2J^*}\right)^2 + \tfrac{H^*}{J^*}}$. Equation (17) shows that the forbidden band is arisen for $D_1^* \neq D_2^*$ because the magnitude of right-hand side is greater than the unity.

Figure 3(a) and 3(b) show spin-wave forbidden (black) and allowed (white) bands as a function of lattice constant $l$ of magnonic crystal and DMI strength $D_2$ in the region 2, respectively, when $D_1 = 0$ mJ/m². We find that the numerically-obtained boundaries between the allowed and forbidden bands, indicated by blue symbols, are in agreement with Eq. (17). The allowed bands become narrower when either the lattice constant or the DMI strength in the region 2 increases because either potential width or potential height increases. We note that the frequencies of the allowed and forbidden bands are located in the sub-THz ranges, which is much higher than gigahertz frequency ranges in ferromagnets. This result suggests that magnonic crystal operated at high frequencies can be constructed using DMI-modulated antiferromagnets.

### B. Easy axis aligned with z axis

In this case, when we ignore the equilibrium magnetization tilting, there is no DMI effect on the spin-wave dispersion because the DM vector ($\hat{\mathbf{y}}$) is orthogonal to the easy axis. Therefore, alternating allowed and forbidden bands, shown in Fig. 3, do not exist in this case. As described in the Sec. II. B., however, two linearly polarized spin waves experience different dispersions due to the magnetization tilting at the DMI step, which in turn results in different phase shifts for two linearly-polarized spin waves [see Eqs. (13) and (14)]. These polarization-dependent phase shift provides a way to construct a magnonic band structure, which is qualitatively and quantitatively different from that shown in the previous section.

We consider a one-dimensional magnonic device having a magnonic crystal, which is connected to homogeneous DMI regions [see Fig. 4(a) for schematic]. In the homogeneous region, we attach a metallic layer (i.e., Pt layer) to measure spin accumulation through the inverse spin Hall effect. Here we consider that the metallic layer covers 1000 magnetic moments. In a low spin memory loss limit, the spin accumulation originating from spin dynamics is given as $\boldsymbol{\mu} = -\hbar \sum_i (\mathbf{S}_i \times \dot{\mathbf{S}}_i)$ [75,76].

The left panel in Fig. 4(b) shows spin-wave forbidden band and allowed band as a function of DMI strength $D_2$ in the region 2. We find that three distinguishable regions are formed in the frequency domain. In the white region, both two linearly-polarized spin waves are allowed, whereas in the black region, both spin waves are forbidden. The grey region corresponds to the polarization-dependent spin-wave band. In this grey region, incident circular-polarized spin waves changes to linearly ($\hat{x}$)-polarized spin waves. Therefore, in this grey region, the DMI step (thus, corresponding magnetization tilting) serves as a spin-wave polarizer.

The right panel in Fig. 4(b) shows the time average of the $z$-component of spin accumulation ($\langle\mu_z\rangle$) as a function of frequency when $D_2 = 1.2$ mJ/m$^2$. We find that $\langle\mu_z\rangle$ oscillates as a function of frequency in the white region of the left panel. This oscillation can be understood as follows. As shown in Fig. 2, a spin wave acquires a frequency-dependent phase shift when passing through a DMI step. Since a magnonic crystal consists of many DMI steps, the frequency-dependent phase shift becomes large. As a result, the output spin accumulation measured at the Pt layer shows a strong oscillation as a function of spin-wave frequency. We also find that $\langle\mu_z\rangle$ is zero in the grey region because this region corresponds to a linearly-polarized spin wave, which carries no angular momentum. This frequency-dependent spin accumulation signal shows that the phase shift induced by the magnetization tilting can be used as another type of magnonic crystal, which allows a spin wave with a specific polarization to pass through the magnonic crystal.

## IV. Summary

In this work, we investigate antiferromagnetic spin-wave propagation in the presence of inhomogeneous DMI. We find that propagating antiferromagnetic spin waves have distinct properties depending on the easy-axis direction because the equilibrium spin torque originating from the inhomogeneous DMI varies with a relative orientation of the equilibrium magnetization with respect to the DM vector. When the easy axis is aligned with the DM vector ($\hat{\mathbf{y}}$), the inhomogeneous DMI affects the boundary condition of spin waves at a DMI step, making the spin-wave amplification possible. Moreover, it also affects the spin-wave dispersion, allowing a magnonic crystal with alternating DMI.

When the easy-axis is orthogonal to the DM vector, there is no DMI effect on the spin-wave dispersion but the equilibrium spin torque emerges, which tilts the staggered order **n** in antiferromagnets. This locally-tilted antiferromagnetic order makes a difference in the spin-wave dispersion for two linearly-polarized spin waves. As a result, a DMI step causes different phase shifts for two linearly-polarized spin waves, which can be used to construct another type of magnonic crystal, based on the spin-wave polarization. Even though not studied here, one can combine these two cases by making the easy axis neither aligned along the DM vector, nor orthogonal to the DM vector. Then both features of allowed and forbidden bands shown in Figs. 3 and 4 coexist.

Our work provides a theoretical tool to investigate the aforementioned rich antiferromagnetic spin-wave physics in the presence of inhomogeneous DMI and would be useful to realize high-frequency spin-wave magnonic crystals based on antiferromagnets.

**ACKNOWLEDGMENTS**

This work was supported by the National Research Foundation of Korea (Grants No. NRF-2017R1A2B2006119) and KU-KIST Graduate School of Converging Science and Technology Program.

# References

[1] V. Baltz, A. Manchon, M. Tsoi, T. Moriyama, T. Ono, and Y. Tserkovnyak, *Rev. Mod. Phys.* **90**, 015005 (2018).

[2] T. Jungwirth, X. Marti, P. Wadley, and J. Wunderlich, *Nat. Nanotechnol.* **11**, 231 (2016).

[3] S.-H. Yang, K.-S. Ryu, and S. Parkin, *Nat. Nanotechnol.* **10**, 221 (2015).

[4] T. Shiino, S.-H. Oh, P. M. Haney, S.-W. Lee, G. Go, B.-G. Park, and K.-J. Lee, *Phys. Rev. Lett.* **117**, 087203 (2016).

[5] O. Gomonay, T. Jungwirth, and J. Sinova, *Phys. Rev. Lett.* **117**, 017202 (2016).

[6] K.-J. Kim et al., *Nat. Mater.* **16**, 1187 (2017).

[7] S.-H. Oh, S. K. Kim, D.-K. Lee, G. Go, K.-J. Kim, T. Ono, Y. Tserkovnyak, and K.-J. Lee, *Phys. Rev. B* **96**, 100407(R) (2017)

[8] S. A. Siddiqui, J. Han, J. T. Finley, C. A. Ross, and L. Liu, *Phys. Rev. Lett.* **121**, 057701 (2018).

[9] L. Caretta et al., *Nat. Nanotechnol.* **13**, 1154 (2018).

[10] S.-H. Oh and K.-J. Lee, *J. Magn.* **23** 196 (2018)

[11] T. Okuno et al., arXiv:1903.03251 (2019)

[12] J. Barker and O. A. Tretiakov, *Phys. Rev. Lett.* **116**, 147203 (2016).

[13] X. Zhang, Y. Zhou, and M. Ezawa, *Sci. Rep.* **6**, 24795 (2016).

[14] Y. Hirata et al., *Nat. Nanotechnol.* **14**, 232 (2019).

[15] C. Kittel, *Phys. Rev.* **82**, 565 (1951).

[16] F. Keffer and C. Kittel, *Phys. Rev.* **85**, 329 (1952).

[17] R. Cheng, D. Xiao, and A. Brataas, *Phys. Rev. Lett.* **116**, 207603 (2016).

[18] R. Khymyn, I. Lisenkov, V. Tiberkevich, B. A. Ivanov, and A. Slavin, *Sci. Rep.* **7**, 43705 (2017).

[19] A. Fischer, H. Zündorf, F. Kaschura, J. Widmer, K. Leo, U. Kraft, and H. Klauk, *Phys. Rev. Applied* **8**, 054012 (2017).

[20] Ø. Johansen and J. Linder, *Sci. Rep.* **6**, 33845 (2016).

[21] D.-K. Lee, B.-G. Park, and K.-J. Lee, *Phys. Rev. Applied* **11**, 054048 (2019).

[22] S. A. Nikitov, P. Tailhades, and C. S. Tsai, *J. Magn. Magn. Mat.* **236**, 320 (2001).

[23] K.-S. Lee, D.-S. Han, and S.-K. Kim, *Phys. Rev. Lett.* **102**, 127202 (2009).

[24] V. V. Kruglyak, S. O. Demokritov, and D. Grundler, *J. Phys. D: Appl. Phys.* **43**, 264001 (2010).

[25] B. Lenk, H. Ulrichs, F. Garbs, and M. Münzenberg, *Phys. Rep.* **507**, 107 (2011).

[26] J. Ding, M. Kostylev, and A. O. Adeyeye, *Phys. Rev. Lett.* **107**, 047205 (2011).

[27] A. V. Chumak, A. A. Serga, and B. Hillebrands, *Nat. Commun.* **5**, 4700 (2014).

[28] K. Vogt, F. Y. Fradin, J. E. Pearson, T. Sebastian, S. D. Bader, B. Hillebrands, A. Hoffmann, and H. Schultheiss, *Nat. Commun.* **5**, 3727 (2014).

[29] A. V. Chumak, V. I. Vasyuchka, A. A. Serga, and B. Hillebrands, *Nat. Phys.* **11**, 453 (2015).

[30] S. Klingler, P. Pirro, T. Brächer, B. Leven, B. Hillebrands, and A. V. Chumak, *Appl. Phys. Lett.* **105**, 152410 (2014).

[31] J. Lan, W. Yu, R. Wu, and J. Xiao, *Phys. Rev. X* **5**, 041049 (2015).

[32] W. Yu, J. Lan, R. Wu, and J. Xiao, *Phys. Rev. B* **94**, 140410(R) (2016).

[33] X. Xing and Y. Zhou, *Npg Asia Materials* **8**, e246 (2016).

[34] J. Stigloher et al., *Phys. Rev. Lett.* **117**, 037204 (2016).

[35] J. Mulkers, B. Van Waeyenberge, and M. V. Milošević, *Phys. Rev. B* **97**, 104422 (2018).

[36] A. A. Serga, A. V. Chumak, and B. Hillebrands, *J. Phys. D: Appl. Phys.* **43**, 264002 (2010).

[37] A. D. Karenowska, J. F. Gregg, V. S. Tiberkevich, A. N. Slavin, A. V. Chumak, A. A. Serga, and B. Hillebrands, *Phys. Rev. Lett.* **108**, 015505 (2012).

[38] A. V. Chumak, V. I. Vasyuchka, A. A. Serga, M. P. Kostylev, V. S. Tiberkevich, and B. Hillebrands, *Phys. Rev. Lett.* **108**, 257207 (2012).

[39] M. Jamali, J. H. Kwon, S.-M. Seo, K.-J. Lee, and H. Yang, *Sci. Rep.* **3**, 3160 (2013).

[40] M. Krawczyk and D. Grundler, *J. Phys.: Cond. Matter* **26**, 123202 (2014).

[41] J. H. Kwon, J. Yoon, P. Deorani, J. M. Lee, J. Sinha, K.-J. Lee, M. Hayashi, and H. Yang, *Sci. Adv.* **2**, e1501892 (2016).

[42] R. Cheng, M. W. Daniels, J. G. Zhu, and D. Xiao, *Sci. Rep.* **6**, 24223 (2016).

[43] J. Lan, W. Yu, and J. Xiao, *Nat. Commun.* **8**, 178 (2017).

[44] I. E. Dzyaloshinskii, *Sov. Phys. JETP* **5**, 1259 (1957).

[45] T. Moriya, *Phys. Rev.* **120**, 91 (1960).

[46] K. Zakeri, Y. Zhang, J. Prokop, T. H. Chuang, N. Sakr, W. X. Tang, and J. Kirschner, *Phys. Rev. Lett.* **104**, 137203 (2010).

[47] D. Cortés-Ortuño and P. Landeros, *J. Phys.: Con, Matter* **25**, 156001 (2013).

[48] J.-H. Moon, S.-M. Seo, K.-J. Lee, K.-W. Kim, J. Ryu, H.-W. Lee, R. D. McMichael, and M. D. Stiles, *Phys. Rev. B* **88**, 184404 (2013).

[49] K. Di, V. L. Zhang, H. S. Lim, S. C. Ng, M. H. Kuok, J. Yu, J. Yoon, X. Qiu, and H. Yang, *Phys. Rev. Lett.* **114**, 047201 (2015).

[50] H. T. Nembach, J. M. Shaw, M. Weiler, E. Jué, and T. J. Silva, *Nat. Phys.* **11**, 825 (2015).

[51] J. Cho et al., *Nat. Commun.* **6**, 7635 (2015).

[52] J. M. Lee, C. Jang, B.-C. Min, S.-W. Lee, K.-J. Lee, and J. Chang, *Nano Lett.* **16**, 62 2016).

[53] J.-V. Kim, R. L. Stamps, and R. E. Camley, *Phys. Rev. Lett.* **117**, 197204 (2016).

[54] R. Cheng, M. W. Daniels, J.-G. Zhu, and D. Xiao, *Sci. Rep.* **6**, 24223 (2016).

[55] S.-J. Lee, J.-H. Moon, H.-W. Lee, and K.-J. Lee, *Phys. Rev. B* **96**, 184433 (2017).

[56] I.-S. Hong, S.-W. Lee, and K.-J. Lee, *Curr. Appl. Phys.* **17**, 1576 (2017).

[57] I.-S. Hong and K.-J. Lee, *Appl. Phys. Lett.* **115**, 072406 (2019).

[58] J. Mulkers, B. Van Waeyenberge, and M. V. Milošević, *Phys. Rev. B* **95**, 144401 (2017).

[59] A. C. Swaving and R. A. Duine, *Phys. Rev. B* **83**, 054428 (2011).

[60] K. M. D. Hals, Y. Tserkovnyak, and A. Brataas, *Phys. Rev. Lett.* **106**, 107206 (2011).

[61] S. K. Kim, Y. Tserkovnyak, and O. Tchernyshyov, *Phys. Rev. B* **90**, 104406 (2014).

[62] K.-W. Kim, H.-W. Lee, K.-J. Lee, and M. D. Stiles, *Phys. Rev. Lett.* **111**, 216601 (2013).

[63] T. Kikuchi, T. Koretsune, R. Arita, and G. Tatara, *Phys. Rev. Lett.* **116**, 247201 (2016).

[64] M. D. Stiles and A. Zangwill, *Phys. Rev. B* **66**, 014407 (2002).

[65] R. F. L. Evans, W. J. Fan, P. Chureemart, T. A. Ostler, M. O. A. Ellis, and R. W. Chantrell, *J. Phys.: Con. Matter* **26**, 103202 (2014).

[66] R. Hertel, W. Wulfhekel, and J. Kirschner, *Phys. Rev. Lett.* **93**, 257202 (2004).

[67] K. Nawaoka, S. Miwa, Y. Shiota, N. Mizuochi, and Y. Suzuki, *Appl. Phys. Express* **8**, 063004 (2015).

[68] W. Zhang et al., *Appl. Phys. Lett*. **113**, 122406 (2018).

[69] J. Liu, M. Shi, J. Lu, and M. P. Anantram, *Phys .Rev. B* **97**, 054416 (2018).

[70] T. Srivastava et al., *Nano Lett.* **18**, 4871 (2018).

[71] G. Chen, T. Ma, A. T. N'Diaye, H. Kwon, C. Won, Y. Wu, and A. K. Schmid, *Nat. Commun.* **4**, 2671 (2013).

[72] J. Torrejon, J. Kim, J. Sinha, S. Mitani, M. Hayashi, M. Yamanouchi, and H. Ohno, *Nat. Commun.* **5**, 4655 (2014).

[73] A. Belabbes, G. Bihlmayer, F. Bechstedt, S. Blügel, and A. Manchon, *Phys. Rev. Lett.* **117**, 247202 (2016).

[74] S. Tacchi, R. E. Troncoso, M. Ahlberg, G. Gubbiotti, M. Madami, J. Akerman, and P. Landeros, *Phys. Rev. Lett.* **118**, 147201 (2017).

[75] R. Cheng, J. Xiao, Q. Niu, and A. Brataas, *Phys. Rev. Lett.* **113**, 057601 (2014).

[76] A. Qaiumzadeh, H. Skarsvåg, C. Holmqvist, and A. Brataas, *Phys. Rev. Lett*. **118**, 137201 (2017).
17

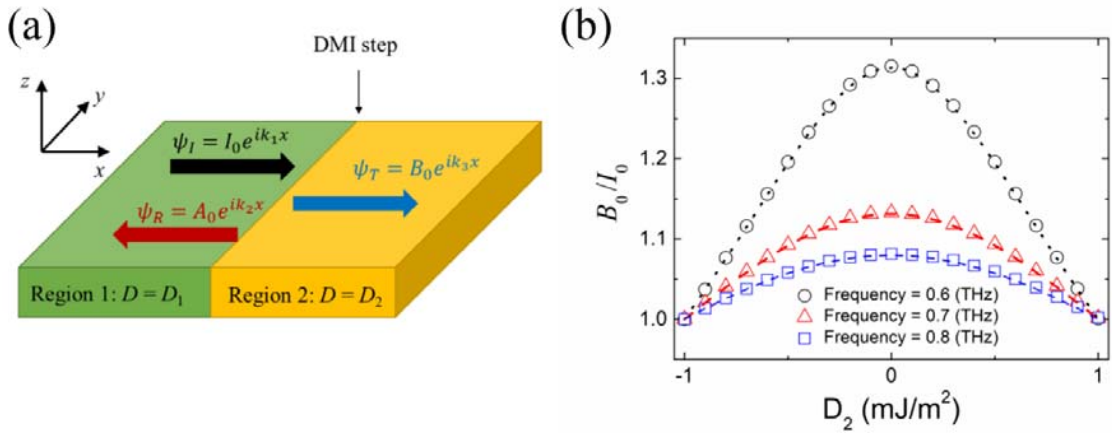

FIG. 1. (a) A schematic illustration of a DMI step. (b) Spin-wave transmittance at a DMI step for $D_1 = -1$ mJ/m$^2$. Symbols are numerical results and dotted lines are Eq. (10).

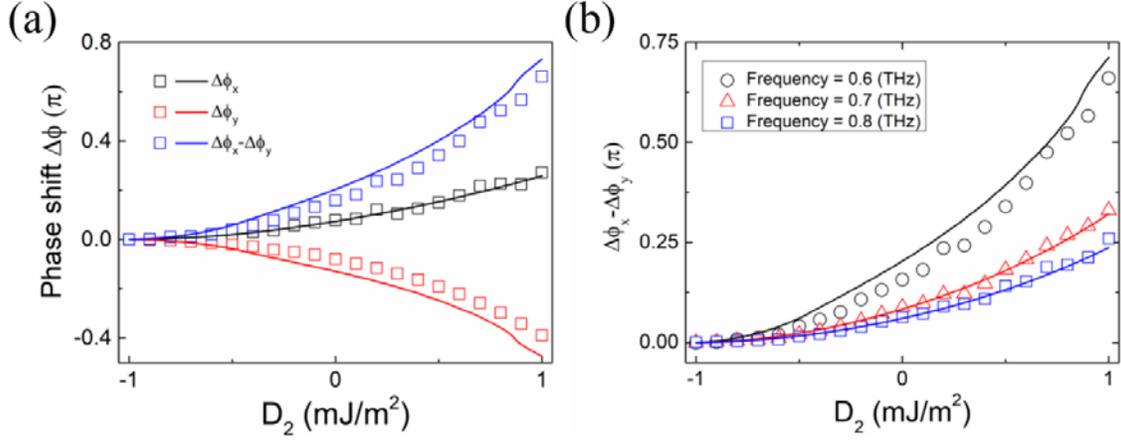

FIG. 2. (a) Phase shift ($\Delta\phi_x$, $\Delta\phi_y$ and $\Delta\phi_x - \Delta\phi_y$) as a function of $D_2$ ($f$ = 0.6 THz). (b) Relative phase shift between two linear spin waves (= $\Delta\phi_x - \Delta\phi_y$) as a function of $D_2$, The $\Delta\phi_x - \Delta\phi_y$ decreases with increasing frequency ($D_1 = -1$ mJ/m$^2$). Symbols are results from numerical calculation of atomistic LLG equation and lines are calculated from Eq. (15).

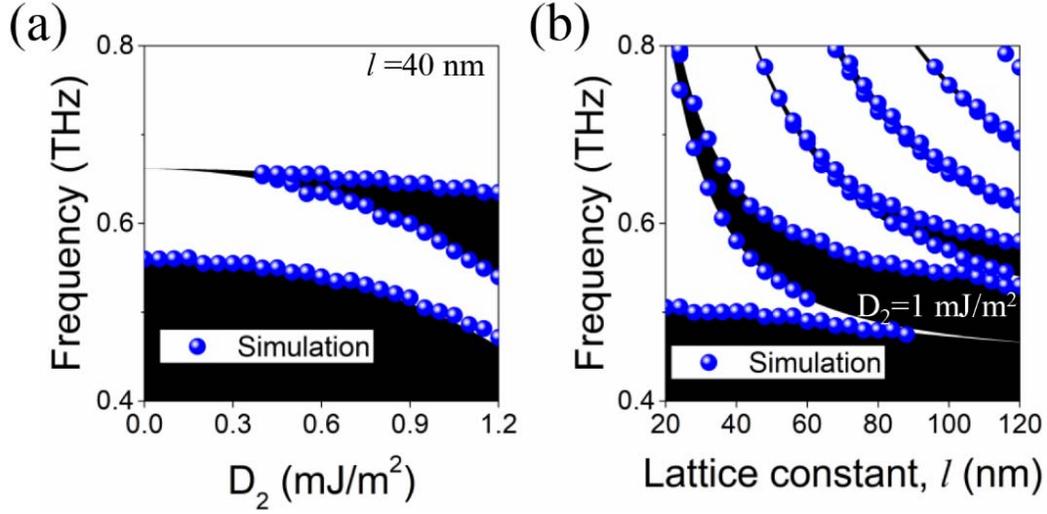

FIG. 3. (a) The spin-wave frequency versus $D_2$ ($l$ = 40 nm, and $D_1$ = 0 mJ/m$^2$). (b) The spin-wave frequency versus the lattice constant $l$ of an alternating DMI region ($D_1$ = 0 mJ/m$^2$, and $D_2$ = 1 mJ/m$^2$). The black (white) regions are forbidden (allowed) bands calculated from analytical dispersion relation in magnonic crystals. Symbols are obtained from numerical simulations.

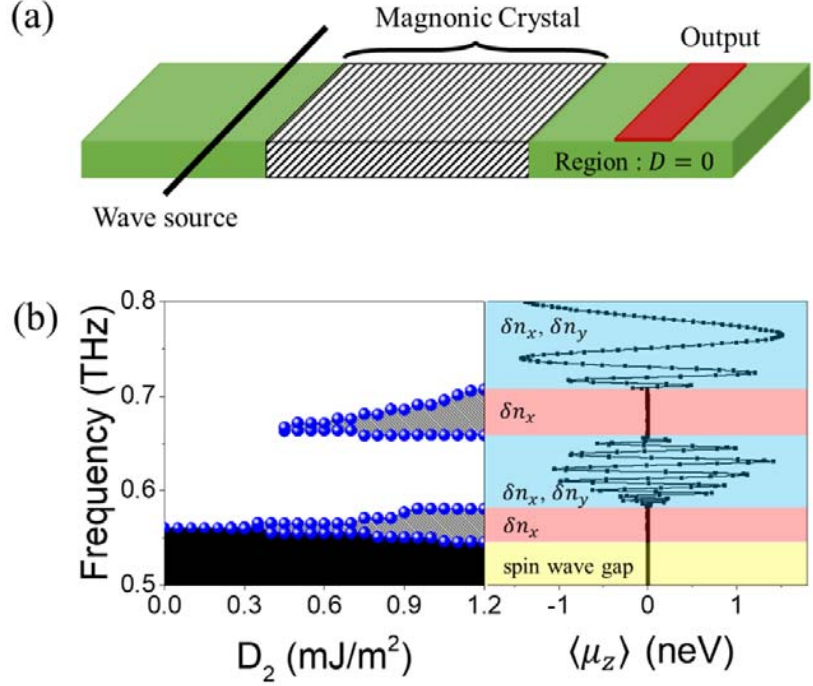

FIG. 4. (a) Schematic diagram for a magnonic crystal based on spin-wave polarization. We assume the DMI of green regions is zero. (b) Left: the spin-wave frequency versus $D_2$ ($l = 40$ nm, and $D_1 = 0$ mJ/m$^2$). Symbols are numerically obtained boundaries between allowed and forbidden bands. Black and white regions correspond to forbidden and allowed bands, respectively. Grey regions are the forbidden bands for $y$-polarized spin wave but the allowed bands for $x$-polarized spin wave. Right: the $z$-component of spin accumulation at the output region in the schematic diagram for $D_2 = 1.2$ mJ/m$^2$. Yellow (blue) regions are forbidden (allowed) spin-wave bands. Red regions correspond to forbidden bands for $y$-polarized spin wave but allowed bands for $x$-polarized spin wave.